\documentclass{elsart}
\usepackage{epsfig}
\usepackage{rotate}
\usepackage{amssymb}
\usepackage{amsmath}

\newcommand\pp{{A'}}
\newcommand\g{{\gamma }}

\newcommand\vdecay{A' \rightarrow e^+  e^-}
\newcommand\xdecay{A' \rightarrow e^+  e^-}
\newcommand\xxdecay{X \rightarrow e^+  e^-}
\newcommand\bra{Br(\eta (\eta') \to \gamma A')}
\newcommand\brx{Br(\eta (\eta') \to \gamma X)}

\newcommand\pair{e^+ e^-}
\newcommand\ee{e^+e^-}

\begin{document}
\begin{frontmatter}
\title{\boldmath Constraints on sub-GeV hidden sector gauge bosons from a search for heavy neutrino decays}
\begin{center}
S.N. Gninenko\\
\vskip0.5cm
{\em Institute for Nuclear Research, Moscow 117312}
\end{center}

\begin{abstract}
Several models of dark matter motivate the concept of hidden sectors consisting of $SU(3)_C \times SU(2)_L \times U(1)_Y$ singlet fields. The interaction between our and hidden matter could be  transmitted by 
new abelian $U'(1)$ gauge  bosons $A'$ mixing with ordinary photons.     
If such $A'$'s with the mass in the sub-GeV range   exist, they 
would be produced through mixing with photons emitted in decays of $\eta$ and $\eta'$ neutral mesons 
 generated  by the high energy proton beam in a neutrino target.
The $A'$'s would then penetrate the downstream shielding and be observed in a neutrino detector via their 
$A' \to \ee $decays. Using bounds from the CHARM neutrino experiment at CERN that searched for an excess of $\ee$ pairs from heavy neutrino decays, the area excluding  the $\gamma - A'$ mixing range   
$10^{-7}\lesssim \epsilon \lesssim 10^{-4}$ for  the  $A'$ mass  region 
$ 1 \lesssim M_{A'} \lesssim 500$ MeV is derived.  
The obtained results are also used to constrain models, where a new gauge boson 
$X$ interacts with quarks and  leptons. New upper limits on  the branching ratio as small as $Br(\eta \to \gamma X) \lesssim 10^{-14}$ and $Br(\eta' \to \gamma X) \lesssim 10^{-12}$  are  obtained, which are several orders of magnitude more restrictive  than the previous bounds from the Crystal Barrel experiment.
\end{abstract}
\begin{keyword} 
hidden sector photons, neutrino decay
\end{keyword}
\end{frontmatter}
The understanding of the origin of dark matter has great importance for cosmology and particle physics.
Several interesting extensions of the Standard Model (SM) dealing with this problem suggest
the existence of `hidden' sectors consisting of $SU(3)_C \times SU(2)_L \times U(1)_Y$
singlet fields. These  sectors   of
particles do not interact with the ordinary matter directly and  couple to 
it by gravity and possibly by other weak forces. For example,  
interaction between our and hidden matter may be  transmitted by a 
new abelian $U'(1)$ gauge  bosons $A'$  (or hidden photons for short) mixing with ordinary photons, 
first discussed by  Okun  in his  paraphoton model  \cite{okun}. If the mixing  strength is very weak or 
the  mass scale of a hidden sector is too high, it is  experimentally unobservable. However, in a class of 
models the $A'$  may have mass and mixing strength lying in the experimentally accessible  and theoretically interesting regions. This makes further searches for $A'$'s  interesting and attractive, for a recent review see \cite{jr}, and references therein.

 In the Lagrangian describing the photon-hidden photon system the only allowed connection between the hidden sector and ours is  given by the kinetic mixing \cite{okun,jr,holdom,foot1}  
\begin{equation}
 L_{int}= -\frac{1}{2}\epsilon F_{\mu\nu}A'^{\mu\nu} 
\label{mixing}
\end{equation}
where  $F^{\mu\nu}$, $A'^{\mu\nu}$ are the ordinary 
 and the  hidden photon  fields, respectively, and $\epsilon$ is their mixing strength.  
In the interesting case when $A'$ has a mass, this kinetic mixing
can be diagonalized resulting in a non-diagonal mass term that 
mixes photons with hidden-sector photons. It means any source of 
$\gamma$'s could produce kinematically possible  massive states $A'$  according to the appropriate mixings. 
Then, when the mass differences are small, photons may oscillate 
into hidden photons, similarly to vacuum neutrino oscillations, with a vacuum mixing angle which is precisely $\epsilon$. If the mass differences are large, it could results in hidden photon decays, e.g. 
into $\ee$ pairs. 

Experimental bounds on the sub-eV and sub-keV  hidden photons
 can be obtained  from searches for an electromagnetic fifth force 
\cite{okun,c1,c2}, from experiments using the method of  photon 
regeneration \cite{phreg,bober,sik,rs,vanb}, and from stellar cooling considerations \cite{seva1,seva2}. For example, it has been pointed out
 that helioscopes searching for solar axions   are sensitive to the keV part of the solar
spectrum of hidden photons and the CAST results \cite{cast1,cast2}
have been translated into limits on the $\g - \pp$ mixing parameter \cite{jr1,jr2,gr,st}.   
Strong bounds on models with additional $A'$ particles  at a low energy scale 
 could be obtained from astrophysical  considerations \cite{blin}-\cite{david}. 
 However, such astrophysical constraints can be relaxed or evaded in 
some models, see e.g. \cite{masso}. New tests on the 
existence of sub-eV hidden photons at new experimental facilities,   such, for example, as SHIPS \cite{ships} or  IAXO \cite{igor} are in preparation.

The $A'$'s in the sub-GeV  mass range, arising in some models, see e.g.
 \cite{prv,bpr,rw,will}, can be explored through the searches for $A'\to \ee$ decays in beam dump experiments \cite{jdb,e137,brun,e141,e774,apex}, or through the rare particle decays, see e.g. \cite{kloe,bes,babar,mami}. For example, if the mass of $A'$ is below   the mass of $\pi^0$, it can be effectively searched for in the decays $\pi^0 \to \gamma A'$, with the  subsequent decay of $A'$ into $\ee$ pair.
  Recently, stringent constraints on the mixing $\epsilon$ in sub-GeV mass range  have been obtained from a search of this  decay mode with existing data of neutrino experiments \cite{sngx,gkx1,nomadx} and from SN1987A cooling \cite{dent}. 
 
It should be noted, that many extensions of the SM  such as  GUTs~\cite{1}, 
super-symmetric~\cite{2}, super-string models~\cite{3} and models including a
new long-range interaction, i.e. the fifth force \cite{carl}, also predict an extra
U$^{'}$(1) factor and therefore the existence of a new gauge boson $X$
corresponding to this new group (we denote  it $X$ to distinguish from $A'$).
The $X$'s could interact directly with quarks and/or leptons, and although
the predictions for its mass are not very firm it could
be light enough ($M_{X}\ll M_{Z}$) for searches  at low energies.
If the mass $M_X$ is in the  sub-GeV range, i.e. of the order of the pion mass,
 an effective search could be conducted for this new vector boson in the radiative decays of neutral  pseudoscalar mesons $P\to \gamma X$, where $P = \pi^{0},\eta$, or 
$\eta^{\prime}$, because  the decay rate of  $P\to \gamma~+~$ 
$\it any~new~particles~with~spin~0~or~\frac{1}{2}$ proves to be negligibly 
small~\cite{di}.\
Therefore, a positive result in the direct search for these decay modes could be
interpreted unambiguously as the discovery of a new light spin 1 particle, in
contrast with other experiments searching for light weakly interacting particles
in rare  $K,~\pi$ or $\mu$ decays~\cite{di,md,gkx2}.
 
Stringent limits on the decay $\pi^0 \to \gamma A'(X)$, $A'(X)\to \ee$, obtained by using results from 
neutrino experiments  have been recently reported in Ref. \cite{sngx}.
The best experimental limits on the branching ratio 
of the decay $\eta (\eta') \to \gamma X$  were obtained by the Crystal Barrel Collaboration  at CERN \cite{amsl1,amsl2}. Using  proton-antiproton annihilation as  a source of  $X$'s from $\eta (\eta')\to \gamma X$ decays, they searched for the corresponding $\gamma$-peak  in their detector, resulting in  90 \% C.L.  upper limits on the 
branching ratio $Br(\eta\to \gamma X) < 6\times 10^{-5}$ for $M_X$  masses  ranging from  200 to  525 MeV,	and	
$Br(\eta' \to \gamma X) < 4\times 10^{-5}$ for $M_X$ between 50 MeV and 925 MeV  \cite{amsl1,amsl2}. 
The goal of this Letter is to show  that  more stringent limits on the decay $\eta (\eta') \to \gamma A'(X)$, followed by the decay $A'(X)\to \ee$, and the mixing 
$\epsilon$ for $A'$ masses up to $\simeq$ 500 MeV  can be 
obtained  from the results of sensitive searches for an excess of single
isolated $\ee$ pairs from decays of heavy neutrinos in the sub-GeV  mass range   by the CHARM  experiment at CERN \cite{charm-1,charm-2}.

The CHARM Collaboration searched for  decays $\nu_h \to \nu \ee$ of heavy neutrinos
in the $\nu_h$ mass range from 10 MeV to 1.8 GeV originated from  decays $\pi, K $ and charmed $D$ mesons decays
\cite{charm-1,charm-2}. The experiment, specifically 
designed to search for neutrino decays in a high-energy neutrino beam,  was performed 
by   using 400 GeV protons 
from the CERN Super Proton Synchrotron (SPS) with the  total number of $2.4\times 10^{18}$ 
protons on (Cu) target (pot). 
The CHARM decay detector (DD), located at the distance of 480 m from the target,
 consist of decay volume of $3\times 3\times 35$ m$^3$ , three  chambers modules located 
inside the volume to detect charged tracks and  followed by a calorimeter.
The decay volume  was essentially  an  empty region to substantially reduced the number of ordinary neutrino interactions. 
The signature of the heavy neutrino decay $\nu_h \to \nu \ee$ were events originating in the decay region at a small angle with respect to the neutrino beam axis with one or two separate electromagnetic showers in the calorimeter 
\cite{charm-2}.
No such events were observed and limits were established on the 
$\nu_{e,\mu} - \nu_h$ mixing strength as a function of the $\nu_h$ mass.
\begin{figure}
 \begin{center}
   \mbox{\hspace{.0cm}\epsfig{file=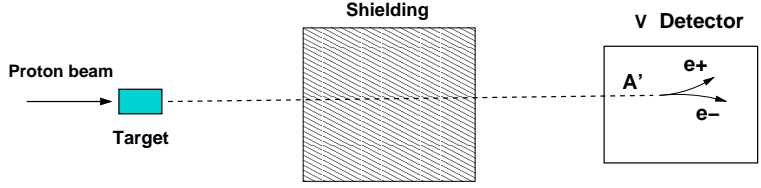,width=100mm}}
\end{center}
    \centering
\vspace{0.5cm}
    \caption{\em Schematic illustration of a proton beam dump  experiment on search for 
$P \to \gamma A', ~šA' \to \ee$ decay chain: neutral mesons $P$ generated by the proton beam in the 
  target produce a flux of high energy $A'$'s through the $\gamma-A'$ mixing in the decay $P\to \gamma \gamma$,  which penetrate the  downstream shielding and decay into $\ee$ pair in a neutrino detector. The same setup can be used to search for  the process $P \to \gamma X, ~šX \to \ee$. See text.}
\label{setup}
\end{figure}
If the  decays $\eta, \eta'\to \gamma A'$ exist, one expects a  flux of 
high energy $A'$'s from the SPS target, since neutral mesons $\eta$ and $\eta'$ are
abundantly produced in the forward direction  by high energy  protons in the target.
If $A'$ is a relatively long-lived particle, this flux  would penetrate the downstream 
shielding without significant attenuation  and would be observed in the CHARM 
detector via the $\xdecay$ decay into a high energy $\ee$ pair, as schematically 
illustrated in Fig. \ref{setup}.
The occurrence of $\xdecay$ decays  would appear as an excess
of $\pair$ pairs in the CHARM DD above those expected from  standard neutrino
interactions. The experimental signature of these events is clean and they can be selected in the CHARM DD with a small background. As  the final states of the  decays $\nu_h\to \nu \ee$ and $\xdecay$ are identical, the  
results of the  searches for the former  can be used  
to constrain the later for the same  $\ee$  invariant mass regions.

The flux of hidden photons from decays of $\eta$'s and $\eta'$'s  produced in the target
by primary protons can be expressed as follows:
\begin{equation}
\Phi(A')\propto N_{pot}\int \frac{d^3\sigma(p + N\to \eta (\eta') + X)}{d^3p_{\eta(\eta')}} 
\epsilon^2 {\rm Br}(\eta (\eta')\to \gamma \gamma)f d^3p_{\eta(\eta')}  
\label{flux}
\end{equation}
where $N_{pot}$ is the  number of pot, $\sigma(p + N\rightarrow \eta (\eta') + X)$
is the $\eta ( \eta')$ meson production cross-section,
$Br(\eta (\eta') \to \gamma \gamma)$ is 
 the $\eta (\eta') \to 2\gamma$ decay mode branching fraction \cite{pdg}, and
$f$ is the decay phase space factor, respectively.

\begin{figure}[htb!]
 \begin{center}
   \mbox{\epsfig{file=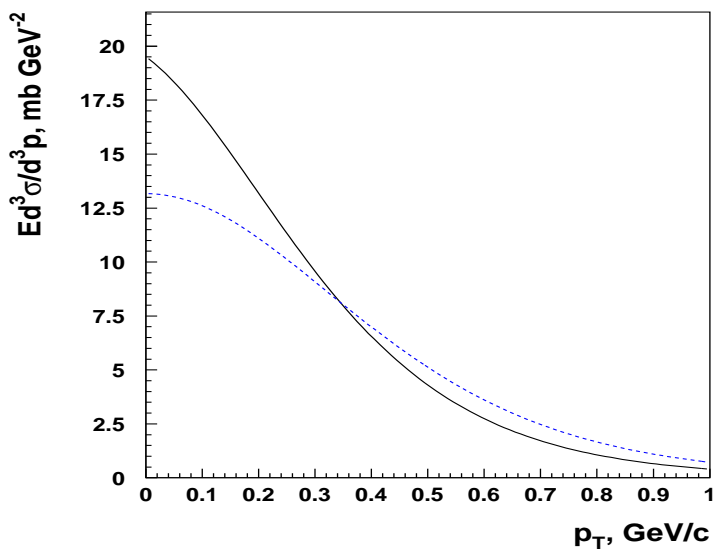, height=80mm,width=100mm}}
 \end{center}
\vspace{0.5cm}
    \centering
    \caption{\em Invariant cross section for $\eta$ mesons production in pp collisions at $\sqrt{s}=27.4$ GeV calculated as a function of $p_T$ in the low-$p_T$ region for the B-G parameterization of Eq.(4) (solid curve), and 
 from Eq.(\ref{lowpt})( dashed curve).  }
 \label{pt}
 \end{figure}

To perform calculations we used simulations of the process shown in Fig.\ref{setup} from our  previous work on 
 $\pi^0\to \gamma A'$ decays \cite{sngx}  by taken into account  
the relative normalization of the yield of  different meson species $\pi^0 : \eta  : \eta'$ from the original 
publications. 
\begin{figure}[htb!]
 \begin{center}
   \mbox{\epsfig{file=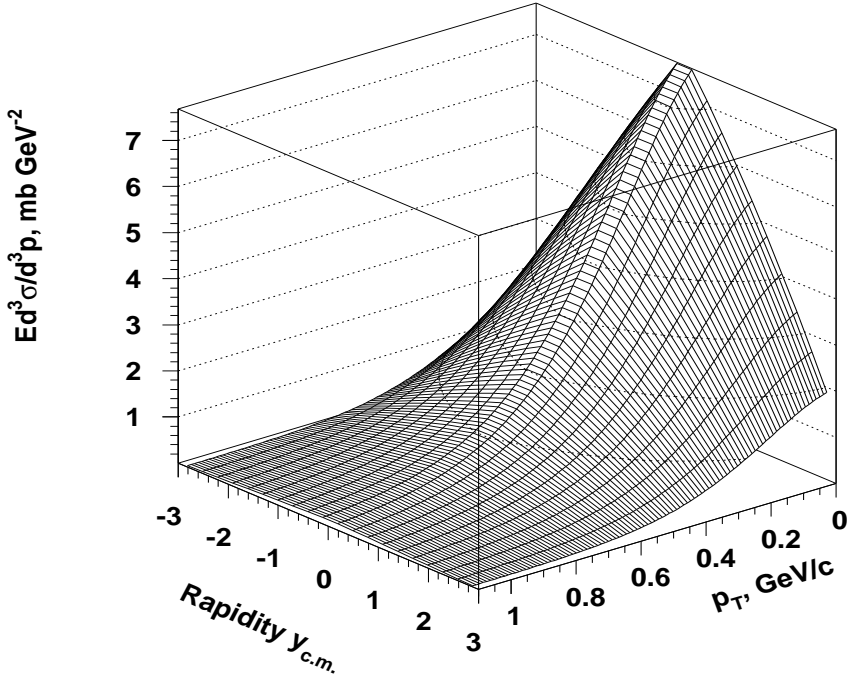, height=100mm,width=140mm}}
 \end{center}
\vspace{0.5cm}
    \centering
    \caption{\em Invariant cross section for $\eta$ production in pp collisions at 
$\sqrt{s}=$ 27.4 GeV  vs $p_T$ and rapidity $y_{c.m.}$ 
over the range $-3.0 < y_{c.m.}< 3.0$. }
 \label{bicrsec}
 \end{figure}
The invariant cross section of a hadron  production  can be expressed as \cite{pdg}
\begin{equation}
E\frac{d^3 \sigma }{d^3p} = \frac{d^3 \sigma }{p_Tdp_T dy d\phi} = \frac{d^2 \sigma }{2\pi p_T dp_T dy}
\label{crsec}
\end{equation}
where $p_T$ is the transverse momentum of the particle, $y$ is its rapidity, and 
 in the last equality integration over the full 2$\pi$
azimuthal angle $\phi$ is performed. For the production cross sections of $\eta$ and $\eta'$ neutral mesons we used the Bourquin-Gaillard (B-G) formula from Ref.\cite{bourq}, which gives the parametric form of 
\eqref{crsec} for the production in high-energy hadronic collisions of many different hadrons over the full phase-space:
\begin{equation}
E\frac{d^3\sigma(p + N\rightarrow \eta (\eta') + X)}{d^3 p}= A_{\eta (\eta')}\bigl(\frac{2}{E_T+2}\bigr)^{12.3} {\rm exp}(-\frac{5.13}{Y^{0.38}}) f(p_T), 
\label{bourq}
\end{equation}
where 
\[f(p_T) = \left\{ 
\begin{array}{l l}
  exp(-p_T) & \quad \mbox{ $p_T<1$ GeV/c}\\
  exp(-1-23(p_T-1)/\sqrt{s}) & \quad \mbox{ $p_T>1$ GeV/c}\\ \end{array} \right. \]
with $Y=y_{max}-y$, being the rapidity, and $E_T(p_T)$ the transverse energy (momentum) in GeV.
Coefficients $A_{\eta (\eta')}$ in (4) are normalization factors that were
tuned to obtain the  cross sections $\sigma_{\eta}$ and  $\sigma_{\eta'}$ in pp collisions 
at 400 GeV/c. In these calculations for the relative yields  of $\pi^0 : \eta  : \eta'$ mesons the  
values $ 1 : 0.078 : 0.024$, respectively, were used, which were obtained from the measurements of the total cross sections $\sigma_{\pi^0}(127.2 \pm 1.5 \pm 3.2$ mb) and  $\sigma_\eta$ ($ 9.78 \pm 0.56$  mb), 
and from an estimate of the $\eta'$ production in pp-interactions at 400 GeV/c by the NA27 experiment at CERN SPS \cite{na27}. 
The invariant $\eta$ production cross section, obtained by using the B-G parameterization (4), is shown 
in Fig. \ref{bicrsec} as a function of $p_T$ and rapidity $y_{c.m.}$ in the center  of mass system.  
The total $\eta$ and $\eta'$  production cross sections in p-Cu collisions were calculated  from its linear extrapolation to the target atomic number.
The B-G $p_T$ distributions at low transverse momenta  were corrected by taking into account the measurements results  of $\pi^0$'s  and $\eta$'s production in 450 GeV p-Be and p-Au collisions down to very low $p_T$ ($\gtrsim 20$ MeV/c) obtained by a  joint experiment of the TAPS and CERES Collaborations  at the CERN SPS \cite{agak}. 
In this experiment  precise measurement data on $\eta$-production at low $p_T$, revealed a lower cross section than expected from extrapolating data at higher $p_T$. The  B-G parameterization, found to be in a good  agreement with 
$\pi^0$ data for the full $p_T$ range, and also with the $\eta$ data for the large ($p_T \gtrsim 0.4$ GeV/c) momentum transfer, overestimates the production rate of $\eta$  mesons by 25-30\% at transverse momenta
 $p_T \lesssim 0.4$ GeV/c.   This observation results in  decrease of the  acceptance of the  CHARM DD for $A'$'s  produced in $\eta (\eta')\to \gamma A'$ decays and, hence has to be  taken into account.  
To  achieve a better description of the low-$p_T$ -region, a parameterization suggested by \cite{agak,alper} was used: 
\begin{equation}
E\frac{d^3 \sigma }{d^3p}= B_{\eta (\eta')} \Bigl(\beta {\rm exp}(-b m_T) + \alpha \frac{(1-x_T)^c}{(1+m^2_T)^4}\Bigr) 
\label{lowpt}
\end{equation}
where $m_T$ is the transverse mass, $x_T = 2m_T/\sqrt{s}$, and parameters $\beta = 0.15, ~ b=6.5~ \rm{GeV^{-1}},~ \alpha = 0.011, ~c=7.9$ and normalization factors $B_{\eta (\eta')}$
were determined in Ref.\cite{agak} by fitting the precise low $p_T$ experimental data in the rapidity range $3 < y < 4$.   The results of fitting were also in good agreement with other meson production data obtained for different rapidity intervals. The comparison of two parameterizations is shown in Fig. \ref{pt}.   We  used Eq.(\ref{lowpt}) also for calculations of low $p_T$ distributions of $\eta'$ mesons.  

The calculated fluxes and energy distributions  of $\eta,\eta'$ produced in the Cu target were used to predict the  
flux of $A'$'s, as a  function of its mass. 
For a given flux $d\Phi(M_{A'}, E_{A'}, N_{pot})/d E_{A'}$ of $A'$'s the expected number of 
$\xdecay$ decays occurring within the fiducial length $L$ of the CHARM 
detector located  at a distance $L'$ from the
neutrino target is given by 
\begin{eqnarray}
N_{\xdecay}=\bra Br(A' \to\ee) \int \frac{d\Phi}{dE_{A'}}\nonumber \\  
\cdot  {\rm exp}\Bigl(-\frac{L'M_{A'}}{P_{A'}\tau_{A'}}\Bigr)\Bigl[1-{\rm exp}\Bigl(-\frac{L M_{A'}}{P_{A'}\tau_{A'}}\Bigr)\Bigr]\zeta A dE_{A'}
\label{nev}
\end{eqnarray}
where $ E_{A'}, P_{A'}$, and $\tau_{A'}$ are the $A'$ energy, momentum and the lifetime  
at rest, respectively, and  $\zeta$ is the $\pair$ pair reconstruction efficiency.
The acceptance $A$ of the DD was calculated  tracing $A'$'s
produced in the Cu-target to the detector taking the
relevant momentum and angular distributions into account. 
As an example for a mass $M_{A'}= 300~\rm MeV$, $A=0.07$ and $\zeta \simeq 0.6$ \cite{charm-1}.
In this estimate the  average momentum is
$<p_{A'}>\simeq 35$ GeV and  $L\simeq 10$ m.
 
\begin{figure}[htb!]
 \begin{center}
   \mbox{\epsfig{file=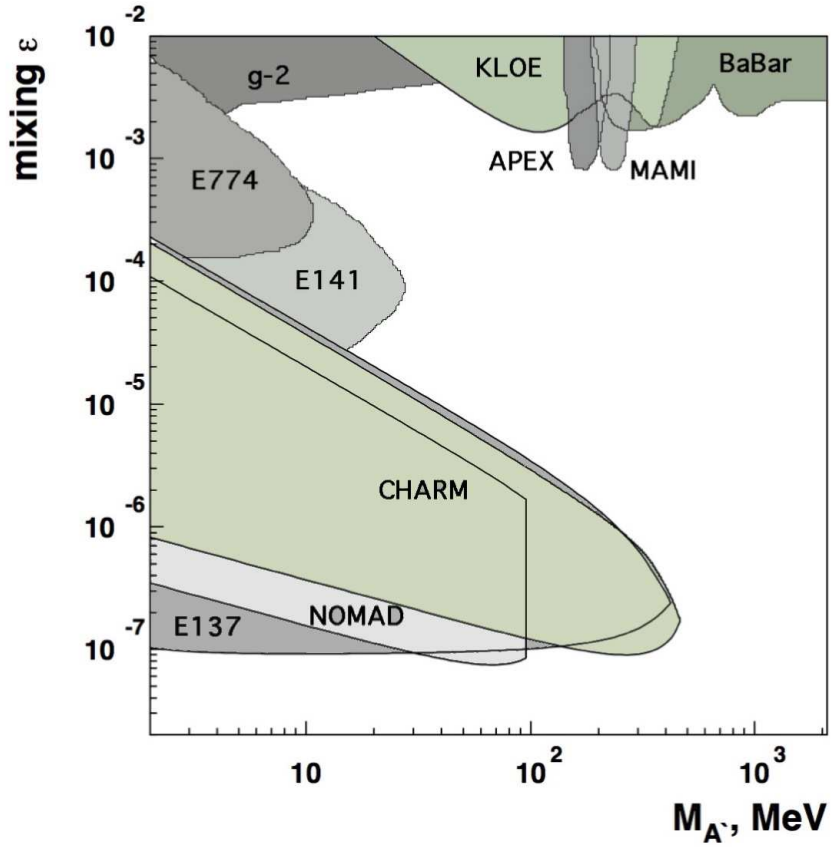, height=120mm,width=120mm}}
 \end{center}
\vspace{0.5cm}
    \centering
    \caption{\em Exclusion region in the ($M_{A'}; \epsilon$) plane 
obtained in the present work from the results of the CHARM experiment \cite{charm-1,charm-2}. 
The areas excluded from the (g-2) considerations, the results  of the Nomad  \cite{sngx} and electron beam dump 
experiments E137 \cite{jdb,e137}, E141 \cite{e141}, E774 \cite{e774},  and from the searches by APEX \cite{apex}, 
KLOE\cite{kloe}, BaBar\cite{babar} and MAMI \cite{mami} are also  shown for comparison.}
 \label{coupl}
 \end{figure}

The obtained results can be used to impose constraints on the previously discussed  models 
with $A'$ hidden photons. For $A'$ masses below   the mass $M_{\eta(\eta')}$ of $\eta(\eta')$ meson, 
the  corresponding branching fraction of  the  decay $\eta(\eta') \to \gamma A'$, is   given by \cite{prv,bpr}: 
\begin{equation}
Br(\eta \to \gamma A') = 2\epsilon^2 Br(\eta \to \gamma \gamma) \Bigl( 1- \frac{M_{A'}^2}{M_{\eta}^2}\Bigr)^3
\label{br}
\end{equation}
 with the similar expression for the $\eta'$ decay.
Assuming  the dominant $A'$ boson decay is into $\ee$ results in a corresponding decay rate,  which  
for small mixing is given by:
\begin{equation}
\Gamma (\vdecay) = \frac{\alpha}{3} \epsilon^2 M_{A'} \sqrt{1-\frac{4m_e^2}{M_{A'}^2}} \Bigl( 1+ \frac{2m_e^2}{M_{A'}^2}\Bigr)
\label{rate}
\end{equation}
Using the relation $N_{\ee}^{90\%} > N_{\xdecay}$, where $N_{\ee}^{90\%}$ (= 2.3 events) is the 90$\%~C.L.$ upper limit for the  number of signal events \cite{charm-1,charm-2} and Eqs.(7,8), one can  determine the $90\%~ C.L.$ exclusion region in the ($M_{A'}; \epsilon $) plane from the results of the  CHARM experiment,
 which is shown in Fig. \ref{coupl} together with regions excluded by the Nomad \cite{sngx} and electron beam dump experiments E137, E141, E774  \cite{jdb,e137,e141,e774},  and by recent measurements from APEX \cite{apex}, KLOE \cite{kloe}, BaBar \cite{babar} and MAMI \cite{mami}. 

The shape of the exclusion contour  from the  CHARM experiment corresponding to the $A'$ mass range
$M_{A'} \gtrsim$ 300 MeV is defined mainly by the phase space factor in \eqref{br}.
Using similar considerations the exclusion area for the decay $\eta' \to \gamma A'$ has been also obtained. 
However, due to the lower $\eta'$ production cross section and a small branching fraction 
of the decay $\eta' \to \gamma \gamma$ this exclusion area  falls
basically within that of obtained for $\eta$ decays. Note, that the  uncertainty for the $\eta$ production at low $p_T$ does not significantly affect the limits derived. For example, the variation of the $\eta$ yield in \eqref{nev}
by 25\% results in the corresponding variation of the limits on mixing strength $\epsilon$  of the order of 5\%. 
This is because the sensitivity of the search is proportional to the $\epsilon^4$. Indeed, in \eqref{nev} the 
branching fraction of \eqref{br} and  the decay  rate $\Gamma (\vdecay)$
of \eqref{rate} both are  proportional to  $\epsilon^2$.

\begin{figure}[htb!]
 \begin{center}
   \mbox{\epsfig{file=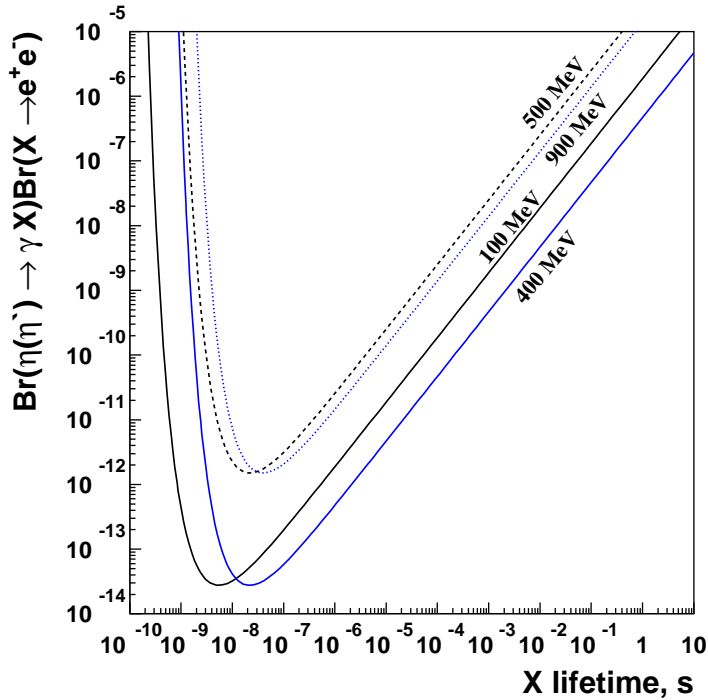, height=100mm,width=100mm}}
 \end{center}
\vspace{0.5cm}
    \centering
    \caption{\em The 90$\%~C.L. $  upper limits on the branching ratio
$Br(\eta(\eta') \to \gamma X) Br(X \to \ee)$ versus $\tau_{X}$ obtained from the CHARM experiment
for $\eta$ (solid curves) and $\eta'$ (dashed curves) decays. The numbers  near the curves indicate the corresponding values of $M_{X}$.}
 \label{limit}
 \end{figure}

We can also constrain models where  the  previously discussed $X$ bosons interact with both quarks and leptons.
The 90$\%~C.L.$ upper limits on the
$\brx Br(\xxdecay)$ vs $X$ lifetime $\tau_{X}$  shown in Fig. \ref{limit} were calculated  by using the relation $N_{\ee}^{90\%} > N_{\xxdecay}$ and  Eq.(\ref{nev}). Our result is sensitive to a branching ratio $Br(\eta (\eta') \to \gamma X) \gtrsim 10^{-14} (10^{-12})$, which is about nine (eight) orders of magnitude smaller than the previous limit from the Crystal Barrel experiment\cite{amsl1,amsl2}. Over most of the $\tau_X$ region, the $X$  lifetime is sufficiently long, that $LM_{X}/p_{X}\tau_{X} \ll L'M_{X}/p_{X}\tau_{X} \ll 1$. 
The interaction of $X$ bosons with both quarks and leptons can be written in the form:
\begin{equation}
L_X= g_X (Q_{BX} B^i+Q_{eX} L_e^i+Q_{\mu X} L_\mu^i+Q_{\tau X} L_\tau^i) X^i
\end{equation} 
where $\alpha_X = g_X^2 /4\pi$ is the coupling constant, $B^i = \sum_{q=u,d,s,..} \overline{q}\gamma^i q $, $L_e^i= \overline{e}\gamma^i e +\overline{\nu}_{eL}\gamma^i \nu_{eL}$, ..., see e.g. \cite{di,gkx2}.  Assuming charges $Q_{BX}\simeq Q_{eX} \simeq 1$,  we found 
\begin{equation}
\alpha_X < k \frac{1}{M_X [MeV]} \Bigl(1- \frac{M_X^2}{M_\eta^2}\Bigr)^{-3/2}, 
\label{alphax}
\end{equation}
which is valid for $k=4.3 \times 10^{-12}$ and $M_X < 2 M_\pi $ ( 270 MeV). For the mass range from 270 to $M_{\eta}$ (548 MeV)
the decay channel $ X\to \pi\pi$ is open. To calculate this decay rate is not a simple task, because  the QCD long distance effects have to be taken into account  \cite{md}.  Therefore,  to avoid this difficulty,  we used for this decay rate  a reasonable estimate   $\Gamma (X\to \pi\pi)   \simeq  \alpha_X Q^2_{BX}M_X$, which results in  $k=1.4 \times 10^{-12}$ for $Q_{BX}\simeq 1$. From the similar considerations, for the  mass range $M_{\eta}<M_X<M_{\eta'}$(948 MeV) the limit on the coupling constant is given by \eqref{alphax} with   replacements $k=3.5 \times 10^{-10}$  and $M_{\eta'}$ instead of $M_{\eta}$.    
 The bounds \eqref{alphax} are valid for $\tau_X \gtrsim 10^{-10} M_{X}$[MeV] s. They are    
 more restrictive than those obtained in \cite{md,gkx2},   and than  bounds reported by NOMAD \cite{nomadx}. Less stringent limits (by a factor $\simeq \alpha$) could be obtained  for the cases where the $X$ interacts only with leptons, or when it is a leptophobic boson which interacts only with quarks  and  decays into $\ee$ pair through the quark loop  \cite{md,gkx2}.
 For $X$ produced in $\eta$ decay with the  masses $ M_{X}<$ 545 MeV/c$^{2}$   the best limits from CHARM  are in the region $Br(\eta\to \gamma X) Br(X\to\ee) \lesssim (2-3)\times 10^{-14}$, while the corresponding limits for $X$'s with 
 masses $ M_{X}<$ 948 MeV/c$^{2}$ from $\eta'$ decays are $Br(\eta'\to \gamma X) Br(X\to\ee) \lesssim 2\times 10^{-12}$.  The attenuation of the $X$- flux due to $X$ interactions with matter was found to be negligible, e.g. for couplings of \eqref{alphax} the $X$ boson mean free
path in iron is much longer, as compared with the iron and earth shielding 
total length of 0.4 km used for the CHARM beam.

 In summary, using sensitive limits from the CHARM experiments on heavy neutrino decays 
$\nu_h \to \nu \ee$, new bounds on a hidden sector gauge  $A'$ bosons 
produced in the neutral meson decay $\eta (\eta') \to \gamma A'$ in the sub-GeV $A'$ mass range are obtained.
The $A'$'s could  mediate interaction between our world and  
 hidden sectors consisting of  $SU(3)_C \times SU(2)_L \times U(1)_Y$ singlet fields through the mixing with 
 ordinary photons. For the $A'$ mass region $ 1 \lesssim M_{A'}\lesssim 500$ MeV  the obtained exclusion area 
 covers  the $\gamma - A'$ mixing strength in the range  $10^{-7}  \lesssim \epsilon \lesssim 10^{-4}$. 
 The results obtained are also applicable to the $\eta$ and $\eta'$ decays into a photon and a new gauge bosons  $X$ 
that interacts with quarks and leptons, or only with quarks. Our best result is sensitive to a branching ratio 
$Br(\eta (\eta') \to \gamma X) \gtrsim 10^{-14}(10^{-12})$, which is about nine (eight) orders of magnitude
stronger than the previous limit from the Crystal Ball experiment\cite{amsl1,amsl2}. 
The obtained constraints were used to set new limits on  the $X$ coupling strength
to lepton and quarks which are more restrictive than bounds from previous searches. 
These results enhance existing motivations for further  more 
sensitive search for $A'(X)$ decay  at the high intensity frontier \cite{hif}
 and additional tests of hidden sectors in neutrino experiments.  

The help of V.D. Samoylenko and D. Sillou in calculations is greatly appreciated.

\end{document}